%% file: main.tex
\documentclass[a4paper]{article}
\usepackage{graphicx}
\usepackage[hidelinks]{hyperref}

\hypersetup{pdftitle={Security assessment of common open source MQTT brokers and clients},pdfauthor={Edoardo Di Paolo, Enrico Bassetti and Angelo Spognardi}}

\begin{document}

\title{Security assessment of common open source MQTT brokers and clients}

\author{Edoardo Di Paolo, Enrico Bassetti, Angelo Spognardi}
\date{\{dipaolo,spognardi,bassetti\}@di.uniroma1.it}

\maketitle

\begin{abstract}
Security and dependability of devices are paramount for the IoT ecosystem. \textit{Message Queuing Telemetry Transport} protocol (MQTT) is the de facto standard and the most common alternative for those limited devices that cannot leverage HTTP. However, the MQTT protocol was designed with no security concern since initially designed for private networks of the oil and gas industry. Since MQTT is widely used for real applications, it is under the lens of the security community, also considering the widespread attacks targeting IoT devices. Following this direction research, in this paper we present an empirical security evaluation of several widespread implementations of MQTT system components, namely five broker libraries and three client libraries. While the results of our research do not capture very critical flaws, there are several scenarios where some libraries do not fully adhere to the standard and leave some margins that could be maliciously exploited and potentially cause system inconsistencies.
\end{abstract}

%------------------------------------------------------------------------------
\input{01-introduction}
\input{02-relatedwork}
%Piccola rassegna degli altri articoli che abbiamo considerato e del perché il nostro è diverso o (se possibile) meglio.
\input{mqtt}

\input{03-methodology}
%L'approccio utilizzato (simile agli altri lavori)

\input{04-experiments}

\section{Conclusions}
%------------------------------------------------------------------------------
MQTT is considered one of the enabling technologies for the IoT ecosystem. It is adopted by almost all IoT applications that run on devices with limited computational power, thanks to the high availability of open source libraries implementing MQTT.
In this paper, we have presented an empirical study of the most popular implementations of both brokers and clients, considering the possible behaviour deviations from the standard that could lead applications to possibly inconsistent or even critical states.
We also tested a physical smart-home device.
The results of our experiments were noticeable: while almost all the considered libraries are correctly handling most of the interactions, as expected, some anomalies have been detected that could be exploited and target the applications, mainly exposing them to denial of service attacks.

Given the promising results, we plan to expand the number of considered libraries further and to include in our experiments some real applications to create some proof-of-concept attacks that exploit the found anomalies. Similarly, we plan to extend the experiments also considering the libraries that also support the new version of the MQTT protocol, namely version 5.
%------------------------------------------------------------------------------
% Refs:
%
\section*{Acknowledgments}
This research has been partially supported  by MIUR (Italian Ministry of Education, University, and Research) under grant ``Dipartimenti di eccellenza 2018-2022'' of the  Computer Science Department of Sapienza University of Rome.

\label{sect:bib}
\bibliographystyle{abbrv}
\bibliography{biblio}

\end{document}

%% file: 01-introduction.tex
\section{Introduction}
\label{sect:introduction}
The number of devices connected to the Internet is growing very
rapidly recently, driving a new wave of technologies and applications
in various fields. One of these trends is the so-called
\textit{Internet-of-Things}: the explosion of low-cost, small/micro
devices (often single-purposes and with an IP stack), Ethernet port and
some space for programming paved the way for a whole new spectrum of
applications.

Usually, these devices have a tiny amount of resources, so that common protocols like HTTP cannot be efficiently implemented without
sacrificing key features of the protocol itself (leading to
non-standard implementation) or key parts of the ``business logic''
(i.e. the main purpose of the device). To overcome this limitation,
several lightweight protocols were invented, like MQTT
(\textit{Message Queuing Telemetry Transport}) protocol or AMQP
(\textit{Advanced Message Queuing Protocol})~\cite{mqtt-amqp}. When
resources are severely limited (simple sensors/actuators) and the system
is in under-constrained environments (low-speed wireless access), the
former is the preferred choice~\cite{mqtt-amqp2}. MQTT is a
publish-subscribe protocol based on a simple message structure, 
basic features and a minimal packet size (considering the message headers). Thanks to this design, nearly all IoT devices use MQTT
or similar lightweight protocols to talk to each other and communicate with the rest of the world. Also, it has undergone several standard processes, and MQTT v. 3.1.1 and 5.0 are both ISO standards ~\cite{iso_2016}.

The protocol was conceived with no security concern, since initially designed for private networks of the oil and gas industry~\cite{mqtt-website}. The adoption of the protocol has ramped up, and several statistics show that many devices use it without any protection~\cite{inproceedings-stats-mqtt}. Also, considering the privacy aspects, given its quite limited features, the MQTT protocol has no built-in encryption features; farther, the use of TLS to provide a secure communication channels is very limited: at the time of writing, comparing with the Shodan search engine the prevalence of the exposed IoT and IIoT devices using MQTT, we have that those that use port 8883 (MQTT over SSL/TLS) is 42, while those using port 1883 (MQTT-unencrypted) is 154632~\cite{inproceedings-stats-mqtt}.
Moreover, MQTT applications keep receiving critiques, with the claim that they adopt weak protocol implementations, even if some of them, like the Mosquitto library, offer extension plugins to
improve security\footnote{\url{https://mosquitto.org/documentation/dynamic-security}} (i.e. role-based authentication or Access Control List, not part of the MQTT standard).

Since MQTT is widely used for real applications, it is under the lens of the security community, also considering the widespread attacks targeting IoT devices. The research is focusing on shifting towards ensuring secure IoT systems, for example, implementing access control mechanisms~\cite{access-control}, lightweight cryptography~\cite{mqttcrypto} or remote attestation of devices~\cite{attestation}. An essential aspect of this context is discovering unforeseen security risks resulting from the necessary interoperability with different implementations of MQTT libraries.

Following this research direction, in this paper we present an empirical security evaluation of several widespread implementations of MQTT system components, namely five broker libraries and three client libraries. Moreover, we also applied our security analysis to an MQTT client embedded in a real IoT device, namely a Shelly DUO Bulb. This IoT device is a remote-controlled LED light bulb. It supports Wi-Fi connectivity and acts as an MQTT subscriber to receive commands, like powering on/off or light dimming.

Our evaluation has aimed to verify the responses of the components of the different libraries to different MQTT messages to see their behaviour in situations where the standard does not indicate clearly how the message (or the connection itself) is supposed to be handled. These mishandling might create interoperability issues or even open doors to malicious attackers. While the results of our research do not capture very critical flaws, there are several scenarios where some of the libraries do not fully adhere to the standard and leave some margins that could be maliciously exploited and potentially cause system inconsistencies.

The structure of the paper is the following: Section~\ref{sect:related-work} reports the state of the art concerning the security analysis of MQTT, while in Section~\ref{sect:methodology} we provide the details about our research methodology. Section~\ref{sect:results} reports the results of our security analysis, and the last section concludes the paper with some remarks and future directions.

%% file: 02-relatedwork.tex
\section{Related works}
\label{sect:related-work}
As the abundance of surveys suggests~\cite{survey,iot,surveymiddle},
security and dependability of IoT devices is paramount for the whole
ecosystem. In this context, the MQTT protocol plays a determinant
role. In 1999 Andy Stanford-Clark (IBM) and Arlen Nipper (then working for
Eurotech, Inc.) proposed the MQTT protocol~\cite{mqtt311} to monitor oil pipelines within the SCADA
framework~\cite{mqtthistory}. Since then, it has been revised in two
main versions, namely 3.1.1 (last update December 2015) and 5 (last
update March 2019). To date, the former is by far the most used in
real applications, the latter being much newer and still not well
adopted~\cite{mqtthistory}.

Like all the network protocols becoming a standard, it has undergone many reviews both formally and empirically.  Several papers focus on MQTT formal modelling and performance analysis~\cite{mqtt-formal-specification,towards-formal,mqtterrors}, others on its possible vulnerabilities, and many others on its security analysis.
In this research, we focused on the security analysis and the comparison of several of the most spread software libraries implementing the MQTT protocol. Instead of using static analysis of their code, as in~\cite{staticanalysis}, we performed a dynamic analysis using the \textit{fuzzing} methodology. In~\cite{novel-fuzzing-approach}, the authors proposed a template-based fuzzing framework and tested its effectiveness against two implementations of MQTT. Using this method, they found some security issues: Moquette and Mosquitto brokers were affected by a vulnerability that would have led to a DoS attack in specific settings if exploited. In our research, we are focusing not only on possible DoS attacks but also on the effects of standard violations of both brokers and clients. Moreover, our analysis applies to five different brokers, three clients and a physical device.

In~\cite{Vaccari2020SlowITeAN}, the authors evaluated the robustness of several MQTT implementations against a subtle family of attacks known as low-rate denial of service. Similarly to this work, a real testbed was set up, and several experiments performed, validating the open vulnerability of all the MQTT implementations.

In~\cite{program-aware-fuzzing-mqtt-applications}, authors describe a new strategy to test MQTT through fuzzing and how much it is efficient against the protocol. However, they do not present any results about the application of their strategy. A similar approach is adopted in~\cite{fuzzing-attacks-for-vuln-discovery}, where the authors propose to apply fuzzing techniques in a container-based environment (Docker). This would allow a large scale test of the MQTT protocol. However, again, the authors do not compare different implementations (they only consider Mosquitto), neither describe the type of attacks they performed.

A different methodology based on \textit{attack patterns}~\cite{automated-security-test} was proposed by \textit{Sochor et al.} and was used to spot hidden vulnerabilities in different broker implementations. They adopted a method to randomly generate test sequences (Randoop) to challenge the different brokers, and they were able to find several failures and unhandled exceptions. Our research adopted a different methodology, tested different broker MQTT implementations, and included clients.

Another methodology to perform a dynamic analysis is model-based testing, as proposed for MQTT applications in~\cite{model-based-testing-mqtt}. The methodology considers using a finite state machine that verifies the properties of the software and proposes extensions to model-based tools for MQTT applications.

% While the previous work focused on stressing the protocol using
% automatic generators for attacks, we decided to tackle the problem by
% looking for issues in the protocol specification itself which can lead
% to implementation issues. This was possible thanks to the fact that
% MQTT is a lightweight protocol.

%%% Local Variables:
%%% mode: latex
%%% TeX-master: "main"
%%% End:

%% file: mqtt.tex
\subsection{MQTT overview}
\begin{figure}[ht]
\centering
\includegraphics[width=0.7\textwidth]{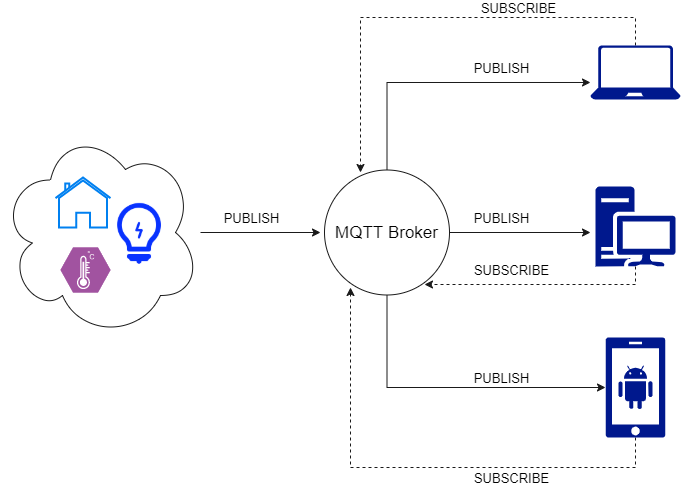}
\caption{Typical MQTT architecture: IoT devices (clients) publish their messages to the broker. Subscribers ask the broker to receive only those messages with topics they subscribe. Broker relays (publishes) to each subscriber only messages with subscribed topics.\label{fig:mqtt}}
\end{figure}

MQTT implements the publish-subscribe communication paradigm (Figure~\ref{fig:mqtt}): \emph{clients} send messages on a \emph{topic} to servers (named \emph{brokers}) that are responsible for delivering them to the interested clients, the final recipients of the messages. Brokers are, then, intermediaries that accept messages and forward a copy of each message to the clients who previously \textit{subscribed} for a given \textit{topic}.  A topic is a UTF-8 string obtained joining one or more topic levels with the slash character, like in \verb|/home/basement/kitchen/temperature/| and the client subscriptions can be made to a topic or part of it, thanks to the use of wildcards, like in \verb|/home/basement/#|.

In a general MQTT session, a client establishes a connection with a broker (CONNECT-CONNACK exchange), subscribes to one or several topics (SUBSCRIBE-SUBACK exchange), publishes contents (PUBLISH-PUBACK exchange), receives other client contents (PUBLISH or PUBLISH-PUBREC-PUBREL exchange, according to the QoS) and terminates the session (client DISCONNECT). The CONNECT packet can implement an authentication mechanism, based on username and password. All the exchanges happen using a clear text TCP session on port 1883 or, if TLS is used, using an encrypted session on port 8883. Encrypted exchanges are mainly used when authentication is enforced, so that username and password are protected against eavesdropping.

In addition to the topic, any message also has a \textit{Quality-of-Service} value (\textit{QoS}), taken in the range 0--2. A QoS equals 0 corresponds to no guarantees, and it means that the message can be lost or delivered multiple times. A client sending a message with QoS equals 1 requires that the message should never be discarded, while it might be delivered multiple times; in this case the sender stores the message until it receives back a PUBACK packet that ensures reception. Similarly, with a QoS set to 2, the message should never be discarded and it should be delivered exactly one time; in this case the client will wait for a PUBREC packet and, once received, it will discard the PUBLISH and it will send a PUBREL packet. The last packet that the client will receive in this exchange is the PUBCOMP that will release the id of the PUBLISH for reuse. It is important to highlight that the publisher QoS is not associated with the subscriber QoS -- unless the implementation supports this non-standardized feature.

%% file: 03-methodology.tex
\section{Methodology}
\label{sect:methodology}

The purpose of our research was to compare the behaviour of different implementations of MQTT. The first step has been finding and setting up the most popular open-source brokers and client libraries that people use to manage their devices or develop common software solutions. To determine the popularity, we took into account the number of stars and forks on GitHub repositories and the number of blog posts citing the examined brokers.

%like the number of projects that use them, or the number of blog posts.
We focused only on open source libraries, namely:
\textit{Mosquitto, EMQ X, HiveMQ, Moquette} and \textit{Aedes} for the
brokers, \textit{paho, mqttools and mqtt.js} for the clients. We will
discuss the brokers and the clients respectively in
Section~\ref{subsect:brokers} and in
Section~\ref{subsect:clients}. Some of these have thousands of
instances running in ``production'' environments, in common consumer
and business-to-business solutions.  We also tested a popular low-cost
\textit{Internet-of-Things} device, namely the Shelly DUO Bulb
(Section~\ref{subsect:physical-device}).

The next step has been to evaluate the type of tests to apply, considering the MQTT standard specification, version 3.1.1. We specifically looked for undefined behaviours, unspecified states or other missing information about message handling. Also, we looked for parts of the standard that might lead to a wrong implementation (e.g. expected actions by the broker/client that are implied but not specified or not clearly specified). This allowed us to focus our testing on a restricted subset of cases, as explained in Section~\ref{sect:results}.

We created different sets of experiments to find possible anomalies in MQTT implementations, developing our fuzzer written in python, with the help of the twisted library\footnote{\url{https://github.com/twisted/twisted}}. Our custom fuzzer allowed us to manage different streams correctly and send custom packets: for example, we could change every bit of the packets to see the brokers behaviour even in the presence of malformed packets.  Standard, common MQTT libraries, instead, implement some state-machine which are expected in some part of the protocol (e.g. QoS2): a straightforward use of such libraries would not allow arbitrary changes in the flow of the messages, like out-of-order messages. Each experiment has been codified in a \textit{JSON} file that specifies the sequence of actions or packets that the test should run on/against the software under test, and the final behaviour of the involved parties have been logged and analyzed.

%%% Local Variables:
%%% mode: latex
%%% TeX-master: "main"
%%% End:

%% file: 04-experiments.tex
\section{Experimental results}

\begin{figure}[ht]
\centering
\includegraphics[width=1\textwidth]{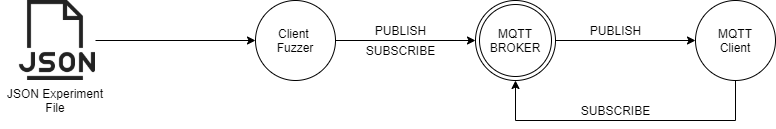}
\caption{Schema of the testbed for the experiments: the fuzzer, which acts as a typical client, takes in input a ``JSON experiment file" containing the client's packets to the MQTT broker. The fuzzer will also receive all the PUBLISH packets sent to the broker. The MQTT Client, instead, uses one of the libraries that are examined in the \autoref{subsect:clients}. \label{fig:testbed}}
\end{figure}

\label{sect:results}
\subsection{Brokers}
\label{subsect:brokers}
\input{brokers}
\subsection{Clients}
\label{subsect:clients}
\input{clients}
\subsection{Physical device}
\label{subsect:physical-device}
\input{physicaldevice}
\subsection{Discussion}
\label{subsect:discussion}
\input{discussion}

%% file: brokers.tex
A broker is a fundamental component in an MQTT architecture. Its job is to accept messages from clients (acting as ``publishers") and then forward them to all clients with subscriptions (``subscribers") matching the topic of the message. This loosely coupled architecture allows the clients to not communicate directly with each other.

Modern brokers support many concurrent connections and messages per second. A flaw in message state machines, packet parsing, topic logic, etc., might expose a high impact vulnerability, which a malicious actor might exploit to launch some attacks like a \textit{Denial-of-Service}.

We analyzed five common MQTT brokers:

\begin{itemize}
    \item \textbf{Mosquitto}\footnote{\url{https://mosquitto.org/}}: it is one of the most used MQTT brokers in the world. It is a single-threaded, lightweight broker written in C. This broker has been widely used thanks to its flexibility.
    \item \textbf{EMQ X}\footnote{\url{https://www.emqx.io/}}: it is written in Erlang and it claims to be so efficient to be ``the Leader in Open Source MQTT Broker for IoT".
    \item \textbf{HiveMQ}\footnote{\url{https://www.hivemq.com/developers/community/}}: a broker written in Java. It supports MQTT version 3.x and 5.0 and it is widely used in automation and industrial systems. We tested the Community Edition.
    \item \textbf{Moquette}\footnote{\url{https://github.com/moquette-io/moquette}}: another Java-powered open-source broker. It is very lightweight but it is less-known and less-used, when compared to other brokers.
    \item \textbf{Aedes}\footnote{\url{https://github.com/moscajs/aedes}}: a broker written in JavaScript/NodeJS. It is the successor of MoscaJS. It does not support version 5 of MQTT, but it is fully compatible with version 3.x and supports several extension libraries.
\end{itemize}

Each broker underwent the same set of tests specified in the next section. We performed more than 60 different experiments on a consumer-grade PC with local connections. A summary of the results is in Table~\ref{tbl:brokersummary}.

\begin{table}[ht]
\caption{Brokers test result summary. The tested versions were the latest stable, available at the time of our experiments.}
\begin{tabular}{|p{2cm}|p{5cm}|p{4cm}|p{1.5cm}|}
\hline
\textbf{Broker} & \textbf{Anomalies found} & \textbf{Security problems} & \textbf{Version}  \\ \hline
Mosquitto & when handling \emph{quality of service}.  & Possible unwanted application scenarios. & 1.16.12 \\ \hline
EMQ X     & when handling \emph{quality of service} and \emph{long topics}.  & Possible unwanted application scenarios. & 4.2.1 \\ \hline
HiveMQ    & when handling \emph{quality of service} and \emph{long topics}.  & Possible unwanted application scenarios. & 2020.5 \\ \hline
Moquette  & when handling \emph{quality of service} and \emph{long topics}. & Possible \emph{denial of service} and unwanted application scenarios. & 0.13 \\ \hline
Aedes     & when handling \emph{quality of service} and \emph{packet references}.  & Possible \emph{denial of service}. & 0.43.0 \\ \hline
\end{tabular}
\label{tbl:brokersummary}
\end{table}

\subsubsection{Experiments and results}

\paragraph{Publish QoS 2 and 1.}
\label{p:qos21aedes}
In this experiment, the client performs the following steps:
\begin{enumerate}
    \item it sends a SUBSCRIBE packet with a specific topic;
    \item it sends the first PUBLISH packet with a \textit{quality of service} 2 and with id 1 over the topic specified in the subscription;
    \item it sends the second PUBLISH packet with a \textit{quality of service} 1, still with id 1 over the topic specified in the subscription;
    \item it sends a PUBREL packet for the first packet sent.
\end{enumerate}
We noticed different broker behaviours: Mosquitto publishes the first received packet with QoS 2, but then it loses the second packet that is not published to the clients due to the PUBCOMP packet that is not received, and so the packet id is not available for use. The EMQ X broker publishes both packets; it handles the flow for the first packet and then the flow for the second one. In HiveMQ and in Moquette, the client that sends packets receives the publication first and after the \textit{pubcomp}, concerning the first packet. Additionally, in HiveMQ the client receives the \textit{pubcomp} back first and then the \textit{pubrec}. Aedes publishes both packets, but the \textit{pubcomp} arrives at the client after the two publications. This behaviour repeated several times, also in the other experiments regarding the \textit{quality of service} that are described below.
%This issue happens several times, also in the next experiments and in other brokers; it is a violation of the MQTT standards as it specifies that the \textit{pubcomp} messages must arrive before the publication of the data to complete the message transmission for the packet with the \textit{QoS} level 2.

\paragraph{Publish QoS 2 and 0.}
This experiment is very similar to the one described above, but in this case, the client performs the following steps:
\begin{enumerate}
    \item it sends a SUBSCRIBE packet with a specific topic;
    \item it sends the first PUBLISH packet with a \textit{quality of service} 2 and with the id 1 over the topic specified in the subscription;
    \item it sends the second PUBLISH packet with a \textit{quality of service} 0 and with the id 1 over the topic specified in the subscription;
    \item it sends a PUBREL packet for the first packet sent.
\end{enumerate}
Mosquitto, in this case, publishes both packets but in reverse order: it handles the one with \textit{quality of service} 0 first, and then it handles, correctly, all the flow regarding the first packet sent with \textit{quality of service} 2. EMQ X and HiveMQ maintain the order of the packets published by the client; also, in the case of HiveMQ, the client received back the \textit{pubcomp} first and then the \textit{pubrec} regarding the packet with \textit{quality of service} 2. Moquette behaves similarly to EMQ X, but, in this case, the \textit{pubcomp} arrives after the publication of the second packet. Aedes has the same behaviour as Mosquitto, but the \textit{pubcomp} arrives after the publication as in the previous experiment.

\paragraph{Double publish QoS 2.}
In this experiment, the client performs the following steps:
\begin{enumerate}
    \item it sends a SUBSCRIBE packet with a specific topic;
    \item it sends the first PUBLISH packet with a \textit{quality of service} 2 and with the id 1 over the topic specified in the subscription;
    \item it sends the second PUBLISH packet with a \textit{quality of service} 2 and with the id 1 over the topic specified in the subscription;
    \item it sends a PUBREL packet for the first packet sent;
    \item it sends a PUBREL packet for the second packet sent.
\end{enumerate}
In Mosquitto, only one publication referred to the first packet sent, but the flow regarding the \textit{quality of service} is properly handled. EMQ X, in this case, has the same behaviour as Mosquitto. Instead, HiveMQ and Moquette publish both packets in the correct order. In Aedes, there is a different behaviour: the broker publishes two packets, but they are the same packet, the first one sent by the client.

\paragraph{Long topic.}
In the MQTT standard, the maximum length topic is 65536 bytes, but we saw that in the source code of EMQ X there is a constant that represents the maximum length, and its value is 4096. So, we tried to subscribe to a topic with more than 4096 bytes. In Mosquitto the subscription to the topic is successful. Instead, HiveMQ cuts the topic to which the client is subscribing to. In Moquette there is an \emph{IOException} and then the client disconnection. In Aedes there is a crash of the broker and the client; in particular, the exception thrown by the experiment generated an error like ``\emph{too many words}''. In EMQ X the client disconnects.

\paragraph{Other experiments.}
Further experiments are listed below. They have been briefly summarized, since the behaviour of all the brokers was correct.
\begin{itemize}
    \item some experiment where the \textit{client id} value in the packet contains characters non-UTF-8 encoded: no anomalies. In detail, we have built a connection packet with the \textit{client id} containing particular characters and the experiment was handled correctly by all brokers;
    \item \textit{Keep-alive} field in connection packet as a string: in all brokers there is the client disconnection due to malformed packet;
    \item subscription (or publication) in an invalid \textit{wildcard}: in all brokers there is the client disconnection due to ``invalid topic'';
    \item topics and \textit{wildcard} encoded in: \textit{utf-16, zzlib, bz2} and \textit{base64}. In the last three cases, there were no anomalies to report. In the \textit{utf-16} experiment, in all brokers the client disconnects. A particular experiment was the one with many `` / '' in the topic value; in \textit{Mosquitto, EMQ X, Moquette} and \textit{Aedes} there was client disconnection while in HiveMQ there was a cut of the topic and then the client subscription;
    \item packets flood with QoS 0: all brokers handled the flood well;
    \item invalid protocol name (or version) in the connection packet: in all cases, the client disconnects;
    \item sending a \textit{pubrel} packet that references a publication packet that was never sent: all brokers, except for Aedes, sent back a \textit{pubcomp} message. In Aedes there is the client disconnection.
\end{itemize}

%% file: clients.tex
In addition to tests on brokers, we also carried out tests on client libraries available in the web. In particular, we studied three different client libraries: \textit{paho-mqtt}\footnote{\url{https://pypi.org/project/paho-mqtt/}}, \textit{mqttools}\footnote{\url{https://pypi.org/project/mqttools/}} and \textit{mqtt.js}\footnote{\url{https://github.com/mqttjs/MQTT.js}}; the first two are written in \textit{python} while the third is written in \textit{javascript}. Again, we considered metrics like the number of stars and forks on GitHub repositories. However, the experiments have not found particular anomalies. Here there is a list of tests we tried:
\begin{itemize}
    \item invalid QoS level: all libraries report an error about the QoS, blocking the sending of the packet;
    \item invalid \textit{wildcard} subscription: in this case \textit{mqtt.js} generates an ``Invalid topic" error, while the other two libraries timeout;
    \item \textit{client id} not encoded in utf-8: in \textit{mqttools} the client cannot connect to the broker, in \textit{paho-mqtt} there is a successful connection to the broker and \textit{mqtt.js} generates an error with the consequent client disconnection;
    \item publication (or subscription) to a topic with a length more than 65536 characters: in all libraries there is the client disconnection.
\end{itemize}

\begin{table}[ht]
\caption{Client libraries test results. The tested versions were the latest stable, available at the time of our experiments.}
\begin{tabular}{|p{2cm}|p{5cm}|p{4cm}|p{1.5cm}|}
\hline
\textbf{Library} & \textbf{Anomalies found} & \textbf{Security problems} & \textbf{Version} \\ \hline
paho-mqtt & when handling subscription (or publication) to an \emph{invalid wildcard}. & It produces an \emph{hang}. & 1.5.1 \\ \hline
MQTT.js   & when handling an invalid \textit{quality of service}. & There is a \emph{crash} of the client due to a \emph{TypeError}. & 4.2.1 \\ \hline %% RIMOSSO DALLA SECONDA CELLA: , for example if we send a packet with an invalid \emph{qos} level, the library generate a \emph{type error}
mqttools  & when handling subscription (or publication) to an \emph{invalid wildcard} and when the \textit{client id} value contains not \emph{utf-8} characters. & It produces an \emph{hang} and an infinite connection loop. & 0.47.0 \\ \hline
\end{tabular}
\label{tbl:clientsummary}
\end{table}

%% file: physicaldevice.tex
In ``home automation'' the MQTT protocol is widely used as most smart devices supports it. Many software applications allow you to use the protocol to manage the smart devices, and one of them, for example, is \textit{Home Assistant}; also \textit{Amazon}, in \textit{AWS IoT}, uses MQTT to connect the user's devices to other devices and other services.

We decided to perform the previous experiments on a physical device. In this case, we tested a \textit{Shelly} light bulb that supports MQTT: it is possible, for example, to turn on or off the device through specific commands sent in the local network. In this device, the protocol configuration can be done in a simple web interface, available in the local network, where the user has to specify the broker's IP and port. The username and password are not mandatory during the configuration; this could be a security issue, depending on the context. This device does not have an ``anti-flood'' regarding the packets it receives; for example, it is possible to turn off and on the light bulb repeatedly and quickly by sending a PUBLISH packet on the specific topic with specific content. The software that runs in the light bulb is the same as other Shelly devices, so this problem also affects them. Therefore, it is possible to send many packets that overload the device's electronic components to make it useless.

To confirm the obtained results in the brokers, we performed on the device the same set of experiments previously carried out on both brokers and clients. For example, the experiment ``Publish QoS 2 and 1" (discussed in Section~\ref{subsect:brokers}) has confirmed the expected results. In this case, we sent a ``turn on packet'' with QoS 2 and then we sent a ``turn off packet'' with QoS 1. When Mosquitto was the broker used, the light bulb turned on but then did not go off, while in all other cases, the light bulb turned on and then turned off.

In addition to these experiments already performed for the various brokers, we have tried to generate some \textit{buffer overflows}, through the payload sent to the device, with a consequent \textit{DoS}. However, the light bulb passed all tests without errors; in particular, the device ignores any form of payload other than what it expects to receive.

%% file: discussion.tex
Interesting results have been obtained from the experiments carried out on brokers, clients, and the physical device. In ~\cite{inproceedings-stats-mqtt} Kant et al. shown that many consumer-grade devices do not use a secure transport for MQTT (like TLS); this is due to the few resources on-board (in turn, this is caused by the target price of these devices in the consumer market, which is very low). Sometimes these devices lack proper authentication protocols, again due to missing resources or insecure default configurations. These security issues could lead attackers to control devices (e.g. they could control an entire \emph{home environment} remotely) directly in some cases (e.g. with \textit{Man-in-the-Middle} attacks).

In our work, the model of the attacker includes the capability to modify the MQTT packet flow, delaying the transmission or making it out of order, or modifying MQTT packets payload, injecting invalid values. This capability can be exploited with limited access to the broker or intermediate network devices, or even remotely, by using other attacks like \textit{Distributed Denial-of-Service} or \textit{flooding} against a network device in the path of the packet flow (for delaying packets, for example). Some of these vulnerabilities can be exploited with an older version of TLS protocol itself\footnote{Note that low-cost IoT devices often implement old protocols, sometimes even partially}: for example, SSL used a vulnerable \textit{Message Authentication Code} until TLS~\cite{dierks1999tls}; vulnerabilities in TLS HMAC implementations are still found years after the standard~\cite{CVE-2015-4458}.

We described bad behaviours for brokers in Section~\ref{p:qos21aedes}. Some of them can be classified as vulnerabilities, and they can lead to attacks if some conditions happen. In fact, we saw that brokers publish some messages violating the protocol state machine in some tests. An example of this is the out-of-order Aedes (and other brokers) use of \textit{pubcomp}, which might be used to trigger a \textbf{replay attack} if the \textit{pubcomp} itself is delayed or dropped by a malicious actor. This attack can disrupt or even damage some devices: for example, IoT-mechanical devices can be continuously triggered until the mechanical part is over-stressed.

Another violation of the standard which leads to a vulnerability (in all brokers but Mosquitto) is the bad handling of long topics: in the MQTT standard, the maximum length topic is 65536 bytes. However, trying to publish to a very long topic ($>$4096 bytes) leads to a disconnection of the client. A malicious actor that can inject (even indirectly, think user-provided information) some characters in the topic may cause a Denial of Service for that client. \textbf{Even worst, in Aedes there is a crash of the broker itself, leading to a \textit{Denial of Service} for the entire MQTT network}.

Some violations of the standard are so misinterpreted that each broker does a different thing: in \textit{Double publish QoS 2} test, nearly all brokers (the only exception is Mosquitto) violates the "unique identifier" feature of MQTT in different ways. This causes no direct impact as a violation, but it can be exploited if some client library shows some bad handling of this situation.

Among all brokers, our tests show that Mosquitto seems to be the strongest one in terms of MQTT standard adoption, and so the safest from a security point of view.

Instead, client libraries have shown only minor issues, many of them relating to encoding errors or long topic subscription issues. Our tests show that they are quite robust, sometimes even better than some brokers.

%% file: main.bbl
\begin{thebibliography}{10}

\bibitem{staticanalysis}
{ Ferrara, Pietro and Mandal, Amit Kr and Cortesi, Agostino and Spoto, Fausto}.
\newblock {Static analysis for discovering IoT vulnerabilities}.
\newblock {\em International Journal on Software Tools for Technology Transfer}, 23:71--88, 2021.

\bibitem{iot}
A.~{Al-Fuqaha}, M.~{Guizani}, M.~{Mohammadi}, M.~{Aledhari}, and M.~{Ayyash}.
\newblock Internet of things: A survey on enabling technologies, protocols, and applications.
\newblock {\em IEEE Communications Surveys Tutorials}, 17(4):2347--2376, 2015.

\bibitem{program-aware-fuzzing-mqtt-applications}
L.~G. Araujo~Rodriguez and D.~Mac\^{e}do~Batista.
\newblock Program-aware fuzzing for mqtt applications.
\newblock In {\em Proceedings of the 29th ACM SIGSOFT International Symposium on Software Testing and Analysis}, ISSTA 2020, page 582–586, New York, NY, USA, 2020. Association for Computing Machinery.

\bibitem{mqtt311}
A.~Banks and R.~Gupta.
\newblock {MQTT} version 3.1.1 plus errata 01, October 2014.
\newblock \url{http://docs.oasis-open.org/mqtt/mqtt/v3.1.1/mqtt-v3.1.1.html}.

\bibitem{fuzzing-attacks-for-vuln-discovery}
G.~Casteur, A.~Aubaret, B.~Blondeau, V.~Clouet, A.~Quemat, V.~Pical, and R.~Zitouni.
\newblock Fuzzing attacks for vulnerability discovery within {MQTT} protocol.
\newblock In {\em 16th International Wireless Communications and Mobile Computing Conference, {IWCMC} 2020, Limassol, Cyprus, June 15-19, 2020}, pages 420--425. {IEEE}, 2020.

\bibitem{mqtterrors}
M.~{Collina}, M.~{Bartolucci}, A.~{Vanelli-Coralli}, and G.~E. {Corazza}.
\newblock Internet of things application layer protocol analysis over error and delay prone links.
\newblock In {\em 2014 7th Advanced Satellite Multimedia Systems Conference and the 13th Signal Processing for Space Communications Workshop (ASMS/SPSC)}, pages 398--404, 2014.

\bibitem{access-control}
P.~Colombo and E.~Ferrari.
\newblock {Access Control Enforcement within MQTT-Based Internet of Things Ecosystems}.
\newblock In {\em Proceedings of the 23nd ACM on Symposium on Access Control Models and Technologies}, SACMAT '18, page 223–234, New York, NY, USA, 2018. Association for Computing Machinery.

\bibitem{mqtthistory}
{Dave Locke, IBM UK Labs}.
\newblock {MQTT Past, Present and Future: 20 Years of MQTT}, 2019.
\newblock \url{https://info.thingstream.io/hubfs/IBM Watson MQTT , MQTT-SN presentation.pdf}, last accessed: 28/02/2021.

\bibitem{dierks1999tls}
T.~Dierks, C.~Allen, et~al.
\newblock The tls protocol version 1.0, 1999.

\bibitem{mqttcrypto}
{Dinculeană, Dan and Cheng, Xiaochun}.
\newblock {Vulnerabilities and Limitations of MQTT Protocol Used between IoT Devices}.
\newblock {\em Applied Sciences}, 9(5), 2019.

\bibitem{attestation}
A.~Francillon, Q.~Nguyen, K.~B. Rasmussen, and G.~Tsudik.
\newblock A minimalist approach to remote attestation.
\newblock In {\em 2014 Design, Automation Test in Europe Conference Exhibition (DATE)}, DATE '14, pages 1--6, Leuven, BEL, 2014.

\bibitem{novel-fuzzing-approach}
S.~Hernández~Ramos, M.~T. Villalba, and R.~Lacuesta.
\newblock {MQTT Security}: A novel fuzzing approach.
\newblock {\em Wireless Communications and Mobile Computing}, 2018, 2018.

\bibitem{towards-formal}
K.~{Hofer-Schmitz} and B.~{Stojanović}.
\newblock Towards formal methods of iot application layer protocols.
\newblock In {\em 2019 12th CMI Conference on Cybersecurity and Privacy (CMI)}, pages 1--6, 2019.

\bibitem{mqtt-formal-specification}
M.~{Houimli}, L.~{Kahloul}, and S.~{Benaoun}.
\newblock Formal specification, verification and evaluation of the mqtt protocol in the internet of things.
\newblock In {\em 2017 International Conference on Mathematics and Information Technology (ICMIT)}, pages 214--221, 2017.

\bibitem{iso_2016}
ISO.
\newblock {ISO/IEC 20922:2016 Information technology -- Message Queuing Telemetry Transport (MQTT) v3.1.1}, Jun 2016.

\bibitem{inproceedings-stats-mqtt}
D.~Kant, A.~Johannsen, and R.~Creutzburg.
\newblock Analysis of iot security risks based on the exposure of the mqtt protocol.
\newblock Technical report, Technische Hochschule Brandenburg, 01 2021.

\bibitem{survey}
J.~{Lin}, W.~{Yu}, N.~{Zhang}, X.~{Yang}, H.~{Zhang}, and W.~{Zhao}.
\newblock A survey on internet of things: Architecture, enabling technologies, security and privacy, and applications.
\newblock {\em IEEE Internet of Things Journal}, 4(5):1125--1142, 2017.

\bibitem{mqtt-amqp2}
J.~E. {Luzuriaga}, M.~{Perez}, P.~{Boronat}, J.~C. {Cano}, C.~{Calafate}, and P.~{Manzoni}.
\newblock A comparative evaluation of amqp and mqtt protocols over unstable and mobile networks.
\newblock In {\em 2015 12th Annual IEEE Consumer Communications and Networking Conference (CCNC)}, pages 931--936, Jan 2015.

\bibitem{CVE-2015-4458}
{CVE}-2015-4458.
\newblock Available from MITRE, {CVE-ID} {CVE}-2015-4458., June~10 2015.

\bibitem{mqtt-website}
{MQTT.org}.
\newblock {Who invented MQTT}, 2021.
\newblock \url{https://mqtt.org/faq/}, last accessed: 28/02/2021.

\bibitem{mqtt-amqp}
N.~{Naik}.
\newblock Choice of effective messaging protocols for iot systems: Mqtt, coap, amqp and http.
\newblock In {\em 2017 IEEE International Systems Engineering Symposium (ISSE)}, pages 1--7, 2017.

\bibitem{surveymiddle}
M.~A. {Razzaque}, M.~{Milojevic-Jevric}, A.~{Palade}, and S.~{Clarke}.
\newblock Middleware for internet of things: A survey.
\newblock {\em IEEE Internet of Things Journal}, 3(1):70--95, 2016.

\bibitem{automated-security-test}
H.~Sochor, F.~Ferrarotti, and R.~Ramler.
\newblock Automated security test generation for mqtt using attack patterns.
\newblock In {\em Proceedings of the 15th International Conference on Availability, Reliability and Security}, ARES '20, pages 1--9, New York, NY, USA, 2020. Association for Computing Machinery.

\bibitem{model-based-testing-mqtt}
K.~Tanabe, Y.~Tanabe, and M.~Hagiya.
\newblock Model-based testing for mqtt applications.
\newblock In M.~Virvou, H.~Nakagawa, and L.~C.~Jain, editors, {\em Knowledge-Based Software Engineering: 2020}, pages 47--59, Cham, 2020. Springer International Publishing.

\bibitem{Vaccari2020SlowITeAN}
I.~Vaccari, M.~Aiello, and E.~Cambiaso.
\newblock Slowite, a novel denial of service attack affecting mqtt.
\newblock {\em Sensors (Basel, Switzerland)}, 20, 2020.

\end{thebibliography}
